\documentstyle[12pt]{article}
\newcommand{\nc}{\newcommand}
\nc{\renc}{\renewcommand}

\nc{\half}{{\textstyle{1\over2}}}

\nc{\etal}{\mbox{\it et al. }}
\nc{\ie}{{\it i.e.}}
\nc{\eg}{{\it e.g.}}

\renc{\thefootnote}{\arabic{footnote}}
\nc{\capt}[1]{{\bf Figure.} {\small\sl #1}}

\nc{\eqs}[2]{\mbox{Eqs.~(\ref{#1},\,\ref{#2})}}
\nc{\eq}[1]{\mbox{Eq.~(\ref{#1})}}

\nc{\figs}[2]{\mbox{Figs.~(\ref{#1},\,\ref{#2})}}
\nc{\fig}[1]{\mbox{Fig~.(\ref{#1})}}

\nc{\tag}[1]{\label{#1} \marginpar{{\footnotesize #1}}}
\nc{\mtag}[1]{\label{#1} \mbox{\marginpar{{\footnotesize #1}}}}
\renc{\baselinestretch}{1.2}
\newlength{\overeqskip}
\newlength{\undereqskip}
\nc{\be}[1]{\begin{equation} \mbox{$\label{#1}$}}
\nc{\bea}[1]{\begin{eqnarray} \mbox{$\label{#1}$}}
\nc{\Section}[2]{\section{#2}\label{#1}}
\nc{\Bibitem}[1]{\bibitem{#1}}
\nc{\Label}[1]{\label{#1}}

\nc{\eea}{\vspace{\undereqskip}\end{eqnarray}}
\nc{\ee}{\vspace{\undereqskip}\end{equation}}
\nc{\bdm}{\begin{displaymath}}
\nc{\edm}{\end{displaymath}}
\nc{\dpsty}{\displaystyle}
\nc{\bc}{\begin{center}}
\nc{\ec}{\end{center}}
\nc{\ba}{\begin{array}}
\nc{\ea}{\end{array}}
\nc{\bab}{\begin{abstract}}
\nc{\eab}{\end{abstract}}
\nc{\btab}{\begin{tabular}}
\nc{\etab}{\end{tabular}}
\nc{\bit}{\begin{itemize}}
\nc{\eit}{\end{itemize}}
\nc{\ben}{\begin{enumerate}}
\nc{\een}{\end{enumerate}}
\nc{\bfig}{\begin{figure}}
\nc{\efig}{\end{figure}}
\nc{\arreq}{&\!=\!&}
\nc{\arrmi}{&\!-\!&}
\nc{\arrpl}{&\!+\!&}
\nc{\arrap}{&\!\!\!\approx\!\!\!&}
\nc{\non}{\nonumber\\*}
\nc{\align}{\!\!\!\!\!\!\!\!&&}

\def\lsim{\; \raise0.3ex\hbox{$<$\kern-0.75em
      \raise-1.1ex\hbox{$\sim$}}\; }
\def\gsim{\; \raise0.3ex\hbox{$>$\kern-0.75em
      \raise-1.1ex\hbox{$\sim$}}\; }
\nc{\DOT}{\hspace{-0.08in}{\bf .}\hspace{0.1in}}
\nc{\Laada}{\hbox {$\sqcap$ \kern -1em $\sqcup$}}
\nc\loota{{\scriptstyle\sqcap\kern-0.55em\hbox{$\scriptstyle\sqcup$}}}
\nc\Loota{{\sqcap\kern-0.65em\hbox{$\sqcup$}}}
\nc\laada{\Loota}
\nc{\qed}{\hskip 3em \hbox{\BOX} \vskip 2ex}

\nc{\real}{{\rm I \! R}}
\nc{\Z}{{\sf Z \!\!\! Z}}
\nc{\complex}{{\rm C\!\!\! {\sf I}\,\,}}
\def\bigid{\leavevmode\hbox{\small1\kern-3.8pt\normalsize1}}
\def\id{\leavevmode\hbox{\small1\kern-3.3pt\normalsize1}}
\nc{\slask}{\!\!\!/}
\nc{\bis}{{\prime\prime}}
\nc{\pa}{\partial}
\nc{\na}{\nabla}
\nc{\ra}{\rangle}
\nc{\la}{\langle}
\nc{\goto}{\rightarrow}
\nc{\swap}{\leftrightarrow}

\nc{\EE}[1]{ \mbox{$\cdot10^{#1}$} }
\nc{\abs}[1]{\left|#1\right|}
\nc{\at}[2]{\left.#1\right|_{#2}}
\nc{\norm}[1]{\|#1\|}
\nc{\abscut}[2]{\Abs{#1}_{\scriptscriptstyle#2}}
\nc{\vek}[1]{{\rm\bf #1}}
\nc{\integral}[2]{\int\limits_{#1}^{#2}}
\nc{\inv}[1]{\frac{1}{#1}}
\nc{\dd}[2]{{{\partial #1}\over{\partial #2}}}
\nc{\ddd}[2]{{{{\partial}^2 #1}\over{\partial {#2}^2}}}
\nc{\dddd}[3]{{{{\partial}^2 #1}\over
        {\partial #2 \partial #3}}}
\nc{\dder}[2]{{{d #1}\over{d #2}}}
\nc{\ddder}[2]{{{d^2 #1}\over{d {#2}^2}}}
\nc{\dddder}[3]{{d^2 #1}\over
        {d #2 d #3}}
\nc{\dx}[1]{d\,^{#1}x}
\nc{\dy}[1]{d\,^{#1}y}
\nc{\dz}[1]{d\,^{#1}z}
\nc{\dl}[1]{\frac{d\,^{#1}l}{(2\pi)^{#1}}}
\nc{\dk}[1]{\frac{d\,^{#1}k}{(2\pi)^{#1}}}
\nc{\dq}[1]{\frac{d\,^{#1}q}{(2\pi)^{#1}}}

\nc{\cc}{\mbox{$c.c.$ }}
\nc{\hc}{\mbox{$h.c.$ }}
\nc{\cf}{cf.\ }
\nc{\erfc}{{\rm erfc}}
\nc{\Tr}{{\rm Tr\,}}
\nc{\tr}{{\rm tr\,}}
\nc{\pol}{{\rm pol}}
\nc{\sign}{{\rm sign}}
\nc{\bfT}{{\bf T }}
\def\eV{{\rm\ eV}}
\def\GeV{{\rm\ GeV}}

\nc{\cA}{{\cal A}}
\nc{\cB}{{\cal B}}
\nc{\cD}{{\cal D}}
\nc{\cE}{{\cal E}}
\nc{\cG}{{\cal G}}
\nc{\cH}{{\cal H}}
\nc{\cL}{{\cal L}}
\nc{\cO}{{\cal O}}
\nc{\cT}{{\cal T}}
\nc{\cN}{{\cal N}}
\nc{\rvac}[1]{|{\cal O}#1\rangle}
\nc{\lvac}[1]{\langle{\cal O}#1|}
\nc{\rvacb}[1]{|{\cal O}_\beta #1\rangle}
\nc{\lvacb}[1]{\langle{\cal O}_\beta #1 |}
\nc{\bb}{\bar{\beta}}
\nc{\bt}{\tilde{\beta}}
\nc{\ctH}{\tilde{\cal H}}
\nc{\chH}{\hat{\cal H}}
\nc{\1}{\aa}
\nc{\2}{\"{a}}
\nc{\3}{\"{o}}
\nc{\4}{\AA}
\nc{\5}{\"{A}}
\nc{\6}{\"{O}}
\nc{\al}{\alpha}
\nc{\g}{\gamma}
\nc{\Del}{\Delta}
\nc{\e}{\epsilon}
\nc{\eps}{\epsilon}
\nc{\lam}{\lambda}
\nc{\om}{\omega}
\nc{\Om}{\Omega}
\nc{\ve}{\varepsilon}
\nc{\mn}{{\mu\nu}}
\nc{\k}{\kappa}
\nc{\vp}{\varphi}

\nc{\advp}[3]{{\it  Adv.\ in\ Phys.\ }{{\bf #1} {(#2)} {#3}}}
\nc{\annp}[3]{{\it  Ann.\ Phys.\ (N.Y.)\ }{{\bf #1} {(#2)} {#3}}}
\nc{\apl}[3]{{\it  Appl. Phys. Lett. }{{\bf #1} {(#2)} {#3}}}
\nc{\apj}[3]{{\it  Ap.\ J.\ }{{\bf #1} {(#2)} {#3}}}
\nc{\apjl}[3]{{\it  Ap.\ J.\ Lett.\ }{{\bf #1} {(#2)} {#3}}}
\nc{\app}[3]{{\it Astropart.\ Phys.\ }{{\bf #1} {(#2)} {#3}}}
\nc{\cmp}[3]{{\it  Comm.\ Math.\ Phys.\ }{{ \bf #1} {(#2)} {#3}}}
\nc{\cqg}[3]{{\it  Class.\ Quant.\ Grav.\ }{{\bf #1} {(#2)} {#3}}}
\nc{\epl}[3]{{\it  Europhys.\ Lett.\ }{{\bf #1} {(#2)} {#3}}}
\nc{\ijmp}[3]{{\it Int.\ J.\ Mod.\ Phys.\ }{{\bf #1} {(#2)} {#3}}}
\nc{\ijtp}[3]{{\it Int.\ J.\ Theor.\ Phys.\ }{{\bf #1} {(#2)} {#3}}}
\nc{\jmp}[3]{{\it  J.\ Math.\ Phys.\ }{{ \bf #1} {(#2)} {#3}}}
\nc{\jpa}[3]{{\it  J.\ Phys.\ A\ }{{\bf #1} {(#2)} {#3}}}
\nc{\jpc}[3]{{\it  J.\ Phys.\ C\ }{{\bf #1} {(#2)} {#3}}}
\nc{\jap}[3]{{\it J.\ Appl.\ Phys.\ }{{\bf #1} {(#2)} {#3}}}
\nc{\jpsj}[3]{{\it J.\ Phys.\ Soc.\ Japan\ }{{\bf #1} {(#2)} {#3}}}
\nc{\lmp}[3]{{\it Lett.\ Math.\ Phys.\ }{{\bf #1} {(#2)} {#3}}}
\nc{\mpl}[3]{{\it  Mod.\ Phys.\ Lett.\ }{{\bf #1} {(#2)} {#3}}}
\nc{\ncim}[3]{{\it  Nuov.\ Cim.\ }{{\bf #1} {(#2)} {#3}}}
\nc{\np}[3]{{\it  Nucl.\ Phys.\ }{{\bf #1} {(#2)} {#3}}}
\nc{\pr}[3]{{\it Phys.\ Rev.\ }{{\bf #1} {(#2)} {#3}}}
\nc{\pra}[3]{{\it  Phys.\ Rev.\ A\ }{{\bf #1} {(#2)} {#3}}}
\nc{\prb}[3]{{\it  Phys.\ Rev.\ B\ }{{{\bf #1} {(#2)} {#3}}}}
\nc{\prc}[3]{{\it  Phys.\ Rev.\ C\ }{{\bf #1} {(#2)} {#3}}}
\nc{\prd}[3]{{\it  Phys.\ Rev.\ D\ }{{\bf #1} {(#2)} {#3}}}
\nc{\prl}[3]{{\it Phys.\ Rev.\ Lett.\ }{{\bf #1} {(#2)} {#3}}}
\nc{\pl}[3]{{\it  Phys.\ Lett.\ }{{\bf #1} {(#2)} {#3}}}
\nc{\prep}[3]{{\it Phys\. Rep.\ }{{\bf #1} {(#2)} {#3}}}
\nc{\prsl}[3]{{\it Proc.\ R.\ Soc.\ London\ }{{\bf #1} {(#2)} {#3}}}
\nc{\ptp}[3]{{\it  Prog.\ Theor.\ Phys.\ }{{\bf #1} {(#2)} {#3}}}
\nc{\ptps}[3]{{\it  Prog\ Theor.\ Phys.\ suppl.\ }{{\bf #1} {(#2)} {#3}}}
\nc{\physa}[3]{{\it  Physica\ A\ }{{\bf #1} {(#2)} {#3}}}
\nc{\physb}[3]{{\it  Physica\ B\ }{{\bf #1} {(#2)} {#3}}}
\nc{\phys}[3]{{\it Physica\ }{{\bf #1} {(#2)} {#3}}}
\nc{\rmp}[3]{{\it  Rev.\ Mod.\ Phys.\ }{{\bf #1} {(#2)} {#3}}}
\nc{\rpp}[3]{{\it Rep.\ Prog.\ Phys.\ }{{\bf #1} {(#2)} {#3}}}
\nc{\sjnp}[3]{{\it Sov.\ J.\ Nucl.\ Phys.\ }{{\bf #1} {(#2)} {#3}}}
\nc{\spjetp}[3]{{\it Sov.\ Phys.\ JETP\ }{{\bf #1} {(#2)} {#3}}}
\nc{\yf}[3]{{\it Yad.\ Fiz.\ }{{\bf #1} {(#2)} {#3}}}
\nc{\zetp}[3]{{\it Zh.\ Eksp.\ Teor.\ Fiz.\  }{{\bf #1}  {(#2)} {#3}}}
\nc{\zp}[3]{{\it Z.\ Phys.\ }{{\bf #1} {(#2)} {#3}}}
\nc{\ibid}[3]{{\sl ibid.\ }{{\bf #1} {#2} {#3}}}
\nc{\rf}[1]{(\ref{#1})}
\nc{\nn}{\nonumber \\*}
\nc{\bfB}{\bf{B}}
\nc{\bfv}{\bf{v}}
\nc{\bfx}{\bf{x}}
\nc{\bfy}{\bf{y}}
\nc{\vx}{\vec{x}}
\nc{\vy}{\vec{y}}
\nc{\oB}{\overline{B}}
\nc{\oI}{\overline{I}}
\nc{\oR}{\overline{R}}
\nc{\rar}{\rightarrow}
\nc{\ti}{\times}
\nc{\slsh}{\hskip-5pt/}
\nc{\sm}{Standard~Model~}
\nc{\MP}{M_{\rm Pl}}
\nc{\tp}{t_{\rm Pl}}
\nc{\ave}{\bar{E}}

\nc{\eff}{{\rm eff}}
\nc{\kk}{\vek{k}}
\nc{\pp}{{\rm p}}
\nc{\ga}{g_{a\gamma}}
\nc{\vv}{\\}
\nc{\eee}{{\bf E}}
\nc{\bbb}{{\bf B}}
\nc{\qcd}{T_{\rm QCD}}
\nc{\G}{\rm \ G}
\begin{document}

{\title{\vskip-2truecm{\hfill {{\small HU-TFT-95-60\\
        }}\vskip 1truecm}
{\bf The paradox of axions surviving primordial magnetic fields}}


{\author{
{\sc Jarkko Ahonen$^{1}$, Kari Enqvist$^{2}$ }\\
{\sl\small Department of Physics, P.O. Box 9,
FIN-00014 University of Helsinki,
Finland} \\
and \\
{\sc Georg Raffelt$^{3}$ }\\
{\sl\small Max-Planck-Institut f\"ur Physik}\vv
{\sl\small F\"ohringer Ring 6, 80805 M\"unchen, Germany}
}
\maketitle
\vspace{2cm}
\begin{abstract}
\noindent
In the presence of primordial magnetic fields the oscillating cosmic axion
field drives an oscillating electric field. The ensuing dissipation
of axions is found to be inversely proportional to the conductivity of
the primordial plasma. This counterintuitive result is essentially
equivalent to ``Zeno's paradox'' or the ``watched-pot effect'' of quantum
mechanics. It implies that the standard predictions of the cosmic axion
density remain unaltered even if primordial magnetic fields are strong.
\end{abstract}
\vfil
\footnoterule
{\small $^1$jtahonen@rock.helsinki.fi};
{\small $^2$enqvist@pcu.helsinki.fi};
{\small $^3$raffelt@mppmu.mpg.de}
\thispagestyle{empty}
\newpage
\setcounter{page}{1}


Besides neutralinos, axions are the only theoretically well-motivated particle
candidate for the ubiquitous cold dark matter that appears to be required in
the standard picture of cosmic structure formation. Primordial axions were
created by the ``misalignment mechanism'' \cite{misalignment} as well as by
the relaxation of the string network which formed at the Peccei-Quinn phase
transition \cite{strings}. In units of the cosmic critical density the relic
axion abundance is found to be $\Omega_a h^2=\xi\; (10^{-5}\eV/m)^{1.175}$,
where $h$ is the present-day Hubble parameter in units of
$100~\rm km\,s^{-1}Mpc^{-1}$ and $m$ the axion mass. The exact value of the
numerical coefficient $\xi=\cO (1)$ is the subject of some debate
\cite{strings}. However, if axions are the dark matter in the galactic halos
implied by astronomical observations, it appears safe to assume that their
mass lies in the range $10^{-5}\eV\lsim m\lsim 10^{-3}\eV$. The current round
of direct search experiments \cite{experiments} for the first time has a
realistic chance of detecting galactic axions at the lower end of this mass
range.

Because of their nonthermal production, axions are essentially born as a Bose
condensate, i.e. a classical, coherent field oscillation of the axion field.
It is obviously important to understand if these oscillations are damped by
dissipation effects which would thermalize, and thus reduce, the cosmic axion
population. For example, it has been shown \cite{thermal} that the
thermalization by interactions with the cosmic plasma is inefficient for
$m\lsim 10^{-1}\eV$, corresponding to values of the Peccei-Quinn scale
$f_a\gsim 10^8\GeV$. (The axion mass and the Peccei-Quinn scale are related by
$m=0.62\eV\; 10^7\GeV/f_a$.)  A proposed ``coherent'' damping mechanism
\cite{XXX} does not seem to be effective in practice \cite{Turner}. The
possibility of resonant axion-photon conversion in a cosmological magnetic
field of order $10^{-9}\G$ has also been studied, with the conclusion that it
yields no significant axion dissipation \cite{japan}.

We presently study another dissipation mechanism which is expected if strong
primordial magnetic fields exist. They may arise during the early cosmic phase
transitions \cite{pt}, and recently it has been shown that magnetic fields
are indeed a stable feature of a second order (electroweak) phase transition
\cite{davies}. Locally the field could be very large. It is limited only by
primordial nucleosynthesis arguments, which imply that
$B\lsim 3\times 10^{10}\G$ at $t\simeq 10^4$~s \cite{dario}. Because flux
conservation implies that $B\sim R^{-2}$ ($R$ is the cosmic scale factor), at
earlier
times the field could have been much stronger. On dimensional grounds, a
typical scale for magnetic field fluctuations should be $B\sim T^2$ so that at
the time of the electroweak phase transition local fields as high as
$10^{24}\G$ could obtain. Depending on how such a large, random magnetic field
scales at large distances, it could be the seed field needed to explain the
observed galactic magnetic fields \cite{poul}. Let us remark that such large
magnetic fields will not facilitate resonant axion-photon conversion in the
early universe because the magnetic field removes the possibility for
degeneracy in the refractive indices of the axion and photon fields.

Magnetic fields would however couple to the cold axions and produce an
oscillating electric field, an effect which is used in the cavity experiments
which search for galactic axion dark matter \cite{experiments}. In the early
universe, such a field would give rise to a bulk velocity of the charge
carriers (electrons, muons, and above the QCD phase transition temperature,
quarks) and hence to a current. This induced current would rapidly dissipate
by thermal collisions in the hot plasma. If the magnetic field is large this
process might dissipate the cosmic axion energy density. Unexpectedly,
however, we find this damping mechanism to be ineffective because the
conductivity of the primordial plasma is {\it too large}. We believe that the
somewhat paradoxical nature of the axion survival story makes it worthwhile to
communicate these results.



In order to derive the equations of motion for axions coupled to the
electromagnetic field with dissipation we start from the Lagrangian
\be{lagrangian}
\cL = \half \partial_\mu a\partial^\mu a-\half m^2a^2+
\half\eee^2-\half \bbb^2+\ga a\, \eee\cdot\bbb,
\ee
where $a$ is the axion field, $m$ its mass,
\be{coupling}
\ga \equiv \frac{\alpha}{2\pi f_a}
\ee
the axion-photon coupling constant, and $f_a$ the Peccei-Quinn scale. Because
of the Nambu-Goldstone nature of axions, the Lagrangian \eq{lagrangian} is
valid only for $a\ll f_a$.

The equations of motion for the coupled axion-photon system that follow from
this Lagrangian have been derived by several authors \cite{eqsmotion}. In our
case they simplify if we assume that the electromagnetic field is dominated by
a large homogeneous primordial magnetic field $\bbb$ and an induced electric
field $\eee$ while we neglect a higher-order induced magnetic field as well as
the thermal radiation fields. If we include the possibility of an electric
current density ${\bf j}$ the coupled equations of motion are found to be
\bea{eqm1}
\dot \eee&=&-\ga\,\bbb\,\dot a-{\bf j}\,, \cr
\ddot a +m^2 a &=&\ga\,\eee\cdot\bbb\,.
\eea
Note that in the literature the terms proportional to $\ga$ are often
presented with an erroneous relative sign.

The only conceivable macroscopic current density ${\bf j}$ is the one induced
by the electric field $\eee$. Assuming a linear response of the medium we may
use Ohm's law ${\bf j}=\sigma\eee$ where $\sigma$ is the conductivity of the
primordial plasma. From \eq{eqm1} it is then evident that $\bbb$ and $\eee$
are parallel so that we may use $E=|\eee|$ and $B=|\bbb|$ instead. This leads
to our final equations of motion
\bea{eqm1x}
\dot E&=&-\beta \dot a-\sigma E, \cr
\ddot a +m^2 a &=&\beta E,
\eea
where $\beta\equiv \ga B$.

The overall behavior of this system is perhaps easiest to understand if one
uses the
vector potential $A$ as a dynamical electric field variable by virtue of
$E=-\dot A$. After a Fourier transformation the equation of motion is
\be{eqm1xx}
\pmatrix{\omega^2+i\sigma\omega&-i\beta\omega\cr
-i\beta\omega&-\omega^2+m^2\cr}\pmatrix{A\cr a\cr}=0,
\ee
revealing that we have to do with two coupled harmonic oscillators of which
one is damped, \ie\ with an axion-photon mixing phenomenon \cite{mixing}.
This equation has solutions only if the determinant of the matrix vanishes,
giving us the dispersion relations.

One obvious solution is a static mode $\omega_1=0$, corresponding to a
constant $A$ and a vanishing $a$, i.e.\ to no electric or axion field at all.
In the absence of dissipation ($\sigma=0$), there is a second static mode
$\omega_2=0$, even in the presence of mixing ($\beta\not=0$). In the presence
of dissipation this mode is purely damped, \ie\ $\omega_2$ is always purely
imaginary. For large dissipation ($\sigma\gg m$ or $\beta$) one finds
$\omega_2=-i\sigma$. The two remaining modes are $\omega_{3,4}=\pm m$ if there
is no magnetic field ($\beta=0$), \ie\ they correspond to the axion field. It
obtains an electric field admixture for $\beta\not=0$.

In order to understand better the behavior of these mixed modes it is useful
to consider a number of approximations. The natural oscillation frequency is
the axion mass $m$. The ``mixing energy'' $\beta$ is proportional to $1/f_a$
and thus to $m$; numerically $\beta=3.65\times10^{-21}\,(B/{\rm G})\,m$. We
only consider magnetic field strengths small enough that always $\beta\ll m$
so that we are in the ``weak mixing limit''. Any deviation from
$\omega_{3,4}=\pm m$ will then be of order $\beta^2$. The shift of the
real part of $\omega_{3,4}$ is $\pm\half\beta^2/m$ for $\sigma=0$, and less
for $\sigma>0$. The imaginary (damping) part of these frequencies is
\be{imaginary}
{\rm Im}(\omega_{3,4})=\half\beta^2\times
\cases{2\sigma/m^2&for $\sigma\ll m$,\cr
1/\sigma&for $\sigma\gg m$.\cr}
\ee
Therefore, if $\sigma\ll m$ the axion field is damped more strongly for an
increasing conductivity as naively expected. In the ``strong damping limit''
$\sigma\gg m$, on the other hand, the actual damping rate of the axion modes
decreases with increasing $\sigma$.

Actually the conductivity of the primordial plasma is huge. In the regime
$m_e\ll T
\ll\qcd$ one finds for an isotropic relativistic electron gas
\cite{con}
\be{cond}
\sigma={\omega_{\rm plas}^2\over 4\pi\sigma_{\rm coll}n_e}\simeq
{T\over 3\pi\alpha},
\ee
where $\omega_{\rm plas}$ is the plasma frequency and $\sigma_{\rm coll}$ the
collision cross section. This result is valid for fields smaller than the
critical field $B_c=m_e^2/e=4.41\times 10^{13}\G$, above which the electrons
cannot be treated as free, and the conductivity \eq{cond} should be multiplied
by a factor $B/B_c$.

In the nonrelativistic regime, relevant for the recombination time, one should
use the conductivity of a nonrelativistic isotropic hydrogen plasma, given by
\be{nr}
\sigma\simeq {(2T)^{3/2}\over 5\pi^{3/2}\alpha m_e^{1/2}}.
\ee
Either way, the conductivity is very large compared with the axion mass so
that we are always in the ``strong damping limit.''

Because the axion field amplitude is dissipated away at the rate
$\Gamma=\beta^2/(2\sigma )$, the rate for the dissipation
of axion number density is that of the squared amplitude,
$\Gamma=\beta^2/\sigma$. This enables us to
check whether this damping mechanism is effective or not in an expanding
universe in the following way.

The damping of axions would be effective only if the dissipation rate for the
axion number density
exceeds the Hubble expansion rate, \ie\ if
damping would occur on a time scale fast compared with the cosmic expansion
time scale. Assuming magnetic flux conservation $\beta$ scales as $1/R^2$ and
hence essentially as $T^2$, while $\sigma$ scales as $T$ in the relativistic
epoch so that $\Gamma\propto T^3$. The expansion rate is roughly
$T^2/m_{\rm Pl}$ (Planck mass $m_{\rm Pl}$) so that axions must be dissipated
early if they are dissipated at all.  One crudely estimates
$\Gamma/H \approx (\beta/m)^2 m^2 m_{\rm Pl} /T^3$. Even if $\beta/m$ were
near unity at the QCD epoch, the factor $m^2 m_{\rm Pl} /\qcd^3$ is
sufficiently below
unity for the axion masses under consideration. Therefore, axions are not
significantly damped by this mechanism.

This conclusion remains valid if one takes the expansion of the universe
explicitly into account, and also if one considers the nonrelativistic epoch
as we shall now demonstrate.
In order to study the explicit solutions of \eq{eqm1} it is useful to take
$\Theta=a/f_a$ as a dynamical variable describing the axion field. With the
initial conditions $\Theta (t_0)=1$ and $E(t_0)=0$, the approximate solution
for $\beta\ll m\ll\sigma$ is
\be{appro}
\Theta (t)\simeq\exp ({-\frac{\beta^2 (t-t_0)}{2 \sigma}})\cos [m(t-t_0)].
\ee
However, this still neglects the expansion of the universe which we shall now
include.

Assuming a flat Robertson-Walker metric with $R(t)$ the dimensionless
scale factor of the
universe, we define
\be{define}
F_{0i}=-RE_i;~~F_{ij}=\epsilon_{ijk}B_kR^2~.
\ee
Including dissipation as before, one may then derive the following
equations of motion:
\bea{eqm2}
\dot{E}=-\frac{gB_{0}\dot{\Theta}}{R^2}-2 H E - \sigma E,\cr
\ddot{\Theta}+3 H \dot{\Theta} + m^2 \Theta =\frac{gEB_{0}}{f_a^2 R^2}.
\eea
Here $g=\alpha /(2\pi )=g_{a\gamma} f_a$
and $H=\dot R/R$ is the Hubble parameter, and we have
made use of
flux conservation which entails $R^2B={\rm const}=B_0$, so that in a
comoving volume the ratio of the magnetic energy density to radiation
density remains fixed.

In this case one cannot perform a simple mode analysis as in the nonexpanding
case. Instead,
eliminating $E$ from \eq{eqm2}, one obtains a third order differential
equation for $\Theta$:
\be{pitka}
\stackrel{\mbox{...}}{\Theta}+(7H+\sigma)
\ddot\Theta+ (m^2+\beta^2R^{-
4}+12H^2+3\dot H+3H\sigma)\dot\Theta
+(4Hm^2+\sigma m^2)\Theta=0
\ee
where we have used $\beta\equiv\ga B_0=g B_0/f_a$. Assuming a radiation
dominated
universe, we may find an approximate solution to \eq{pitka} by writing
$\Theta(t)=C(t)\Theta_0(t)$, where $\Theta_0(t)$ is the solution in
the absence of expansion, and $C(t)$ is a slowly varying function.
Note that $\sigma\gg m\gg H$ except exactly at the QCD phase transition,
above which the axion mass is actually smaller than $H$. The mass starts to
turn on very
rapidly, however, so that the hierarchy of scales is valid when $T\lsim \qcd$.
Writing $\sigma=\sigma_0(t_0/t)^{1/2}$,
to lowest order in $\beta$ and $H=1/(2t)$ we then obtain
\be{result}
\Theta(t)\simeq\left({t_0\over t}\right)^{3/4}\exp\left[
{\beta^2t_0\over \sigma_0}\left( ({t_0\over t})^{1/2}-1\right)\right]
\cos[m(t-t_0)]~.
\ee

One readily observes that damping cannot compete with the expansion
of the universe, which redshifts the energy of the background
magnetic field, and at $t\to\infty$ damping tends to an asymptotic
value.
Thus the decrease in the axion field density due to magnetic field
catalysed dissipation in the early universe is completely negligible.
For example, for a critical
field $B\simeq 10^{13}\G$ one finds, substituting \eq{cond} into \eq{result},
that dissipation reduces $\Theta$ by a factor $(1-5)\times 10^{-22}$.
This result will remain qualitatively unchanged even if we allow for
magnetic fields greater than $B_c$.

In the nonrelativistic regime the conductivity is given by
\eq{nr} and $H=2/(3t)$, whence \eq{result} is
replaced by
\be{result32}
\Theta(t)\simeq\left({t_0\over t}\right)\exp\left[
{3\beta^2t_0\over 4\sigma_0}\left( ({t_0\over t})^{2/3}-1\right)\right]
\cos[m(t-t_0)]~.
\ee
Therefore, damping effects are again negligible.

The standard predictions of the cosmic axion density remain unaltered even if
there exist strong primordial magnetic fields. The reason is that the electric
field never has a chance to grow large because the ohmic dissipation is so
effective. The pump turning the axion energy into electric energy produces a
mere trickle.

This rather counterintuitive result is closely related to ``Zeno's paradox''
or the ``watched-pot effect'' \cite{Stodolsky}.
If two quantum states mix, the transition rate
between them is suppressed if one of them is repeatedly measured, causing the
system to remain ``frozen'' in the original state. In our
case axions would oscillate into photons because the two fields are coupled by
the external magnetic field. The dissipation of the electric field energy can
be viewed on the quantum level as a photon absorption and thus as a
``measurement'' of the system to be in the electromagnetic state. If it is
measured too frequently, it stays frozen in the initial axion state: the
transition rate is suppressed inversely proportional to the rate of
measurement.

\section*{Acknowledgements}

K.E.\ wishes to thank Vesa Ruuskanen for an illuminating discussion on
ohmic dissipation and J.A.\ Martti Havukainen for informative chats on
numerical algorithms. G.R.\ thanks Leo Stodolsky for discussions about Zeno's
paradox.
This research was supported, in part, by the European Union contract
CHRX-CT93-0120 and in Munich by the Deutsche Forschungsgemeinschaft
grant SFB~375.



\newpage


\end{document}